\documentclass[aps,prl,twocolumn,groupedaddress,superscriptaddress]{revtex4-1}
\usepackage{graphicx}
\usepackage{amssymb}
\usepackage{epstopdf}
\usepackage{amsmath}
\usepackage{longtable}
\usepackage{amsmath}
\usepackage{amsfonts}
\usepackage[colorlinks=true,citecolor=blue,linkcolor=red]{hyperref}
\usepackage{graphicx}
\usepackage{bbold}     
\usepackage[dvipsnames]{xcolor}

\usepackage{array,booktabs,ragged2e}
\usepackage{soul}      
\usepackage{cancel}      
\usepackage{tabularx}     
\usepackage{multirow}
\usepackage{hhline}
\usepackage{titlesec}

\begin{document}


\title{Extremely large magnetoresistance in topologically trivial semimetal $\alpha$-WP$_2$}


\author{Jianhua Du}
\affiliation{Department of Physics, Zhejiang University, Hangzhou 310027, China}
\author{Zhefeng Lou}
\affiliation{Department of Physics, Zhejiang University, Hangzhou 310027, China}
\author{ShengNan Zhang}
\affiliation{Institute of Physics, \'{E}cole Polytechnique F\'{e}d\'{e}rale de Lausanne (EPFL), CH-1015 Lausanne, Switzerland}
\affiliation{National Centre for Computational Design and Discovery of Novel Materials MARVEL, \'{E}cole Polytechnique F\'{e}d\'{e}rale de Lausanne (EPFL), CH-1015 Lausanne, Switzerland}
\author{Yuxing Zhou}
\affiliation{Department of Physics, Zhejiang University, Hangzhou 310027, China}
\author{Binjie Xu}
\affiliation{Department of Physics, Zhejiang University, Hangzhou 310027, China}
\author{Qin Chen}
\affiliation{Department of Physics, Zhejiang University, Hangzhou 310027, China}
\author{Yanqing Tang}
\affiliation{Department of Physics, Zhejiang University, Hangzhou 310027, China}
\author{Shuijin Chen}
\affiliation{Department of Physics, Zhejiang University, Hangzhou 310027, China}
\author{Huancheng Chen}
\affiliation{Department of Physics, Zhejiang University, Hangzhou 310027, China}
\author{Qinqing Zhu}
\affiliation{Department of Physics, Hangzhou Normal University, Hangzhou 310036, China}
\author{Hangdong Wang}
\affiliation{Department of Physics, Hangzhou Normal University, Hangzhou 310036, China}
\affiliation{Department of Physics, Zhejiang University, Hangzhou 310027, China}
\author{Jinhu Yang}
\affiliation{Department of Physics, Hangzhou Normal University, Hangzhou 310036, China}
\author{QuanSheng Wu}
\affiliation{Institute of Physics, \'{E}cole Polytechnique F\'{e}d\'{e}rale de Lausanne (EPFL), CH-1015 Lausanne, Switzerland}
\affiliation{National Centre for Computational Design and Discovery of Novel Materials MARVEL, \'{E}cole Polytechnique F\'{e}d\'{e}rale de Lausanne (EPFL), CH-1015 Lausanne, Switzerland}
\author{Oleg V. Yazyev}
\affiliation{Institute of Physics, \'{E}cole Polytechnique F\'{e}d\'{e}rale de Lausanne (EPFL), CH-1015 Lausanne, Switzerland}
\affiliation{National Centre for Computational Design and Discovery of Novel Materials MARVEL, \'{E}cole Polytechnique F\'{e}d\'{e}rale de Lausanne (EPFL), CH-1015 Lausanne, Switzerland}
\author{Minghu Fang}\email{Corresponding author: mhfang@zju.edu.cn}
\affiliation{Department of Physics, Zhejiang University, Hangzhou 310027, China}
\affiliation{Collaborative Innovation Center of Advanced Microstructure, Nanjing 210093, China}

\date{\today}

\begin{abstract}
Extremely large magnetoresistance (XMR) was recently discovered in many non-magnetic materials, while its underlying mechanism remains poorly understood due to the complex electronic structure of these materials. Here, we report an investigation of the $\alpha$-phase WP$_2$, a topologically trivial semimetal with monoclinic crystal structure (C2/m), which
contrasts to the recently discovered robust type-II Weyl semimetal phase in $\beta$-WP$_2$.
We found that $\alpha$-WP$_2$ exhibits almost all the characteristics of XMR materials: the near-quadratic field dependence of MR, a field-induced up-turn in resistivity following by a plateau at low temperature, which can be understood by the compensation effect, and high mobility of carriers confirmed by our Hall effect measurements. It was also found that the normalized MRs under different magnetic fields has the same temperature dependence in $\alpha$-WP$_2$, the Kohler scaling law can describe the MR data in a wide temperature range, and there is no obvious change in the anisotropic parameter $\gamma$ value with temperature. The resistance polar diagram has a peanut shape when field is rotated in $\textit{ac}$ plane, which can be understood by the anisotropy of Fermi surface. These results indicate that both field-induced-gap and temperature-induced Lifshitz transition are not the origin of up-turn in resistivity in the $\alpha$-WP$_2$ semimetal. Our findings establish $\alpha$-WP$_2$ as a new reference material for exploring the XMR phenomena.
\end{abstract}

\pacs{}

\maketitle


 \section{\textbf{I. INTRODUCTION}}
The giant magnetoresistance (GMR) in multilayers involving ferromagnetic metals \cite{PhysRevLett.61.2472,PhysRevB.39.4828} and colossal magnetoresistance (CMR) in manganese oxide materials \cite{PhysRevB.51.14103,Moritomo1996,Ramirez1997} phenomena have opened a new domain of applications as magnetic memories \cite{Daughton1999,rao1996giant}, magnetic valves \cite{wolf2001spintronics}, as magnetic sensors or magnetic switches \cite{lenz1990review,jankowski2011hall}. In the past several decades, the search for new large MR materials has been one of the most important topics in condensed matter physics and material science. Recently, XMR has been discovered in many non-magnetic materials, such as Dirac semimetals Na$_3$Bi and Cd$_3$As$_2$ \cite{wang2012dirac,wang2013three,xiong2015evidence,liang2015ultrahigh,he2014quantum}, Weyl semimetals of TaAs family \cite{Shekhar2015,Ghimire2015,Huang2015,Du2016,Wang2016,huang2015observation,lv2015experimental,xu2015discovery,shekhar2015extremely}, nodal semimetals ZrSiX (X = S, Se, Te) \cite{singha2017large,ali2016butterfly,wang2016evidence,hu2016evidence,hu2017nearly}, LnX (Ln = La, Y, Nd, Ce; X = Sb, Bi) with simple rock salt structure \cite{tafti2016resistivity,sun2016large,kumar2016observation,alidoust2016new,yu2017magnetoresistance,wakeham2016large,pavlosiuk2016giant}, a class of transition metal dipnictides TmPn$_2$ (Tm = Ta, Nb; Pn = P, As, Sb) \cite{wang2016topological,wang2014anisotropic,shen2016fermi,wu2016giant,xu2016electronic,wang2016resistivity,li2016resistivity,luo2016anomalous,yuan2016large}, and the type-II Weyl semimetals WTe$_2$\cite{soluyanov2015type}, $\beta$-MoP$_2$ and $\beta$-WP$_2$ \cite{autes2016robust,wang2017large,schonemann2017fermi,rundqvist1963x,bzduvsek2016nodal}. XMR is a ubiquitous phenomenon in these seemingly unrelated materials, however, the underlying mechanism of XMR is not completely understood. The near-quadratic field dependence of MR exists in most of these materials, all these materials exhibit a field-induced up-turn in resistivity followed by a plateau at low temperatures.

In general, the MR of a material reflects the dynamics of charge carriers and the topology of the Fermi surface (FS).
Several mechanisms have been proposed to explain the XMR in nonmagnetic semimetals \cite{ziman1972principles,parish2003non}. One is the classical two-band model, which predicts parabolic field dependence of MR in a compensated semimetal, and suggest that a small difference of the electron and holes densities will cause the MR eventually saturated at higher magnetic field, such as in Bi \cite{BismuthPR} and graphite\cite{iye1982high}. However, the MR in WTe$_2$ and NbSb$_2$ does not saturate up to 60~T and 32~T \cite{ali2014large,wang2014anisotropic}, respectively, while in TaAs$_2$ \cite{yuan2016large} it saturates under 45~T at 4.2~K. Another open problem is to understand the linear-field-dependence of MR in Dirac and Weyl semimetals as a quantum effect near the crossing point of the conduction and valence bands, having a linear energy dispersion when the magnetic field is beyond the quantum limit \cite{abrikosov1998quantum,abrikosov2000quantum}. In fact, the rich electronic structure near the Fermi level $E_F$ as well as the spin texture driven by the spin-orbit coupling observed by angle-resolved photoemission spectroscopy (ARPES) may be play an important role in XMR of WTe$_2$ \cite{jiang2015signature,wu2015temperature}. On the other hand, ARPES experiments has confirmed that MoAs$_2$ has a relatively simple bulk band structure with a trivial massless surface state along $\bar{\varGamma}$-$\bar{X}$, and its Fermi surfaces (FSs) dominated by an open-orbit topology rather than closed pockets were suggested to be the origin of the near quadratic XMR in this material \cite{lou2017observation}. From the recent studies of XMR in these non-magnetic compounds, it is clear that the mechanism underlying XMR can be different from compound to compound. Searching for new semimetals with XMR and different electronic structures will help understanding this complexity.

Here, we report a novel non-magnetic semimetal $\alpha$-WP$_2$, which belongs to a group of transition metal dipnictides TmPn$_{2}$ crystalizing in OsGe$_{2}$-type structure \cite{hulliger1964new}. Both time-reversal ($\mathcal{T}$) and inversion ($\mathcal{P}$) symmetries are present, which contrasts this material to noncentrosymmetric $\beta$-WP$_2$ predicted to be a robust type-II Weyl semimetal \cite{autes2016robust}. Our band structure calculations show that $\alpha$-WP$_2$ is a type-II nodal line semimetal if spin-orbit coupling (SOC) is neglected, while it is a topological trivial semimetal when SOC is taken into account. Our magnetotransport measurements reveal that $\alpha$-WP$_2$ exhibits almost all the characteristics of XMR materials: the near-quadratic field dependence of MR, a field-induced up-turn in resistivity following by a plateau at low temperature. Our Hall resistivity measurements demonstrate that $\alpha$-WP$_2$ is a compensated semimetal with high mobility of charge carriers. It was also found that the normalized MRs under different magnetic fields has the same temperature dependence, and the Kohler scaling law can describe the MR data in a wide temperature range. The observed MR exhibits anisotropy upon rotating magnetic field in the $ac$-plane, and there is no obvious change in the anisotropic parameter $\gamma$ value with temperature. Our findings reveal that $\alpha$-WP$_2$ is a new platform for exploring XMR phenomena.\\
 \section{\textbf{II. EXPERIMENT AND CALCULATION METHODS}}
  \subsection{\textbf{A. Crystal growth and magnetotransport measurements}}  The single crystals of $\alpha$-WP$_2$ were grown by the chemical vapor transport method. Raw materials were mixed  and ground into a fine powder, sealed in an evacuated quartz tube with 5 mg/cm$^3$ iodine as a transport agent, then heated to 950 $^{\circ}$C for 2 weeks in a two-zone furnace with a temperature gradient of 100 $^{\circ}$C. Polyhedral crystals were obtained at the cold end of tube. The W: P = 33.7: 66.3 composition was confirmed using the Energy Dispersive X-ray Spectrometer (EDXS). The crystal structure was determined by Single-Crystal X-Ray diffractometer (Rigaku Gemini A Ultra). Electrical resistivity in magnetic field ($H$) and Hall resistivity measurements were carried out by using a Quantum Design Physical Property Measurement System (PPMS).
  \subsection{\textbf{\textbf{B. DFT calculations}}}
Density functional theory (DFT) calculations were carried out using the Vienna \textit{ab} \textit{initio} simulation package (VASP) \cite{KRESSE199615}, \cite{PhysRevB.54.11169}, \cite{PhysRevB.59.1758} with generalized gradient approximation (GGA) of Perdew, Burke and Ernzerhof (PBE) \cite{PhysRevLett.77.3865} for the exchange correlation potential was chosen. A cutoff energy of 360 eV and a $10\times10\times6$ k-point mesh were used to perform the bulk calculations. The nodal-line search and Fermi surface calculations were performed using the open-source software WannierTools\cite{wu2017wanniertools} that is based on the Wannier tight-binding model (WTBM) constructed using Wannier90 \cite{MOSTOFI20142309}.
\begin{figure*}[!htbp]
\includegraphics[width= 16cm]{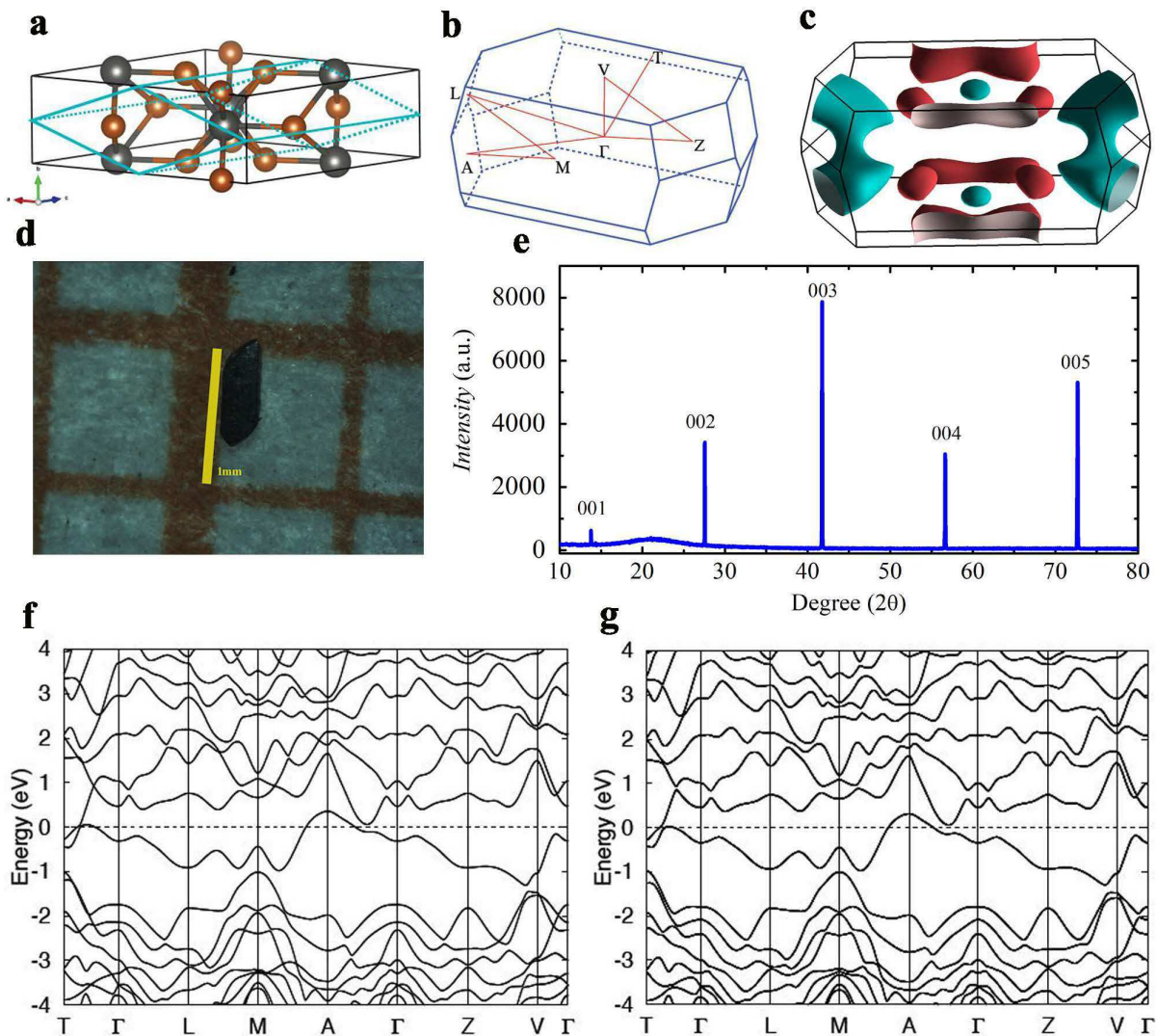}
\caption{(Color online) Crystal structure and calculated band structures of $\alpha$-WP$_2$. (a) Crystal structure of $\alpha$-WP$_2$. (b) The Brillouin zone and (c) the calculated Fermi surface of $\alpha$-WP$_2$. (d) Photograph of a $\alpha$-WP$_2$ crystal. (e) XRD pattern of a $\alpha$-WP$_2$ single crystal. (f,g) Band structures calculated along the high-symmetry path shown in panel (b) without considering SOC (f) and and taking it into account (g).}
 \end{figure*}

 \section{\textbf{III. RESULTS AND DISCUSSIONS}}
  \subsection{\textbf{\textbf{A. Crystal structure and bulk band structures}}}
 Figure 1a shows the crystal structure of $\alpha$-WP$_2$. There is only one position of W atoms and two positions of P atoms in each unit cell, each W atom having eight P atoms as the nearest neighbors.  $\alpha$-WP$_2$ single crystals were grown by a chemical vapor transport method as describing in the Method section. Single crystals with typical dimensions of $0.7\times0.2\times0.1$ mm$^{3}$ were obtained, as shown in Fig. 1d, with (010) being a easy growth axis. Single-crystal X-ray diffraction (XRD) confirmed the monoclinic structure of $\alpha$-phase WP$_2$ with lattice parameters $\textit{a}$ = 8.490(1) \r{A}, $\textit{b}$ = 3.1615(3) \r{A} and $\textit{c}$ = 7.456(1) \r{A}. Figure 1e shows the XRD pattern of a $\alpha$-WP$_2$ crystal.

Based on the above structure and lattice parameters, we carried out density function theory (DFT) calculations as described in the Methods section.
The Brillouin zone (BZ) of the primitive cell is presented in Fig. 1b. The calculated Fermi surface shown in Fig.~1c consists of six electron- (red) and four hole- (cyan) pockets. The electron pockets are closed while hole pockets are connected implying an open-orbit character of the FS. Figure~1f shows the band structure without considering spin-orbit coupling (SOC). A tilted Dirac cone can be seen between \textit{T} and $\varGamma$ which is a part of nodal line, indicating that $\alpha$-WP$_2$ is a type-II nodal line semimetal (NLSM) protected by the $\mathcal{P}$ and $\mathcal{T}$ symmetries when SOC is ignored (more details in the Appendix\cite{appendix}). When SOC is taken into account in calculations, the band-crossing is fully gapped and making $\alpha$-WP$_2$ a semimetal with no band degeneracies as shown in Fig.~1g. Then, we calculate the $\mathbb{Z}$$_{2}$ indices due to the continuous gap in the band structure. This allows to calculate the $\mathbb{Z}$$_{2}$ indices from the parities of occupied wave functions at time-reversal invariant momenta (TRIM) points\cite{LiangFuPRL}. The resulting $\mathbb{Z}$$_{2}$ classification (0;000) identifies $\alpha$-WP$_2$ as a topologically trivial semimetal.

\begin{figure}[!htbp]
    \includegraphics[width= 8cm]{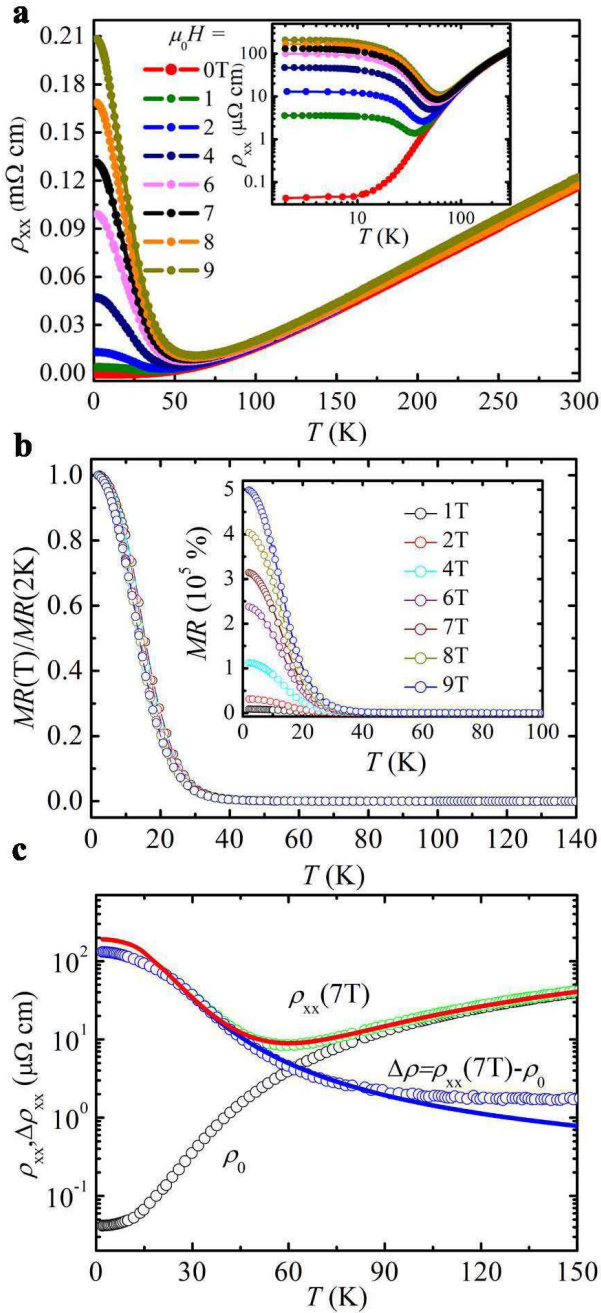}
 \caption{(Color online) Temperature dependence of resistivity of $\alpha$-WP$_2$. (a) Resistivity $\rho$$_{xx}$ of an $\alpha$-WP$_2$ crystal sample as a function of temperature at various magnetic fields. The inset plots the same data on a \textit{log} scale, thus showing the plateaus at lower temperatures. (b) Temperature dependence of the MR normalized by its value at 2 K at various magnetic fields. The inset is the original data of MR as a function of temperature. (c) Temperature dependence of resistivity at
0 T and 7 T, as well as their difference. }
 \end{figure}
  \subsection{\textbf{\textbf{B. Resistivity}}}
 Figure 2a displays the temperature dependence of longitudinal resistivity, $\rho$$_{xx}$($\textit{T}$), under varying magnetic field \textit{\textrm{H}} applied along the direction perpendicular to the \textit{ab} plane as shown in Fig. 1d, with current along the \textit{b} axis. At $\mu_{0}H$ = 0 \textrm{T}, the resistivity decreases monotonically upon decreasing temperature, with the room temperature resistivity $\rho$$_{xx}$(300 K)= 11.75 m$\Omega$~cm and a resistivity $\rho$$_{xx}$(2 K)= 41.74 n$\Omega$ cm at 2~K, the residual resistivity ratio RRR = 2491 of Sample~1 indicating high quality of this $\alpha$-WP$_2$ crystal. Similar to many other XMR materials such as graphite \cite{kopelevich2003reentrant,du2005metal}, bismuth \cite{yang1999large}, PtSn$_{4}$ \cite{mun2012magnetic}, PdCoO$_{2}$ \cite{takatsu2013extremely}, NbSb$_{2}$ \cite{wang2014anisotropic}, and TaP \cite{Du2016}, an up-turn of $\rho_{xx}(T)$ curves under applied magnetic field occurs at low temperatures: the resistivity increases with decreasing temperature, and then saturates, as shown in the inset of Fig.~2a. Intuitively, it seems to be the consequence of a field-induced metal-insulator (MI) transition, as discussed in Refs.~\onlinecite{zhao2015anisotropic,xiang2015multiple,luo2015hall,khveshchenko2001magnetic} predicting an excitonic gap $\Delta$ at low temperature that can be induced by a magnetic field in the linear spectrum of Coulomb interacting quasiparticles. However, the normalized MR, shown in Fig.~2b, has the same temperature dependence  at different magnetic fields, excluding the existence of a magnetic-field-dependent gap. The saturation of $\rho(T, H)$ at low temperatures demonstrates that no gap opening takes place, too. A similar behavior was also observed in WTe$_2$ \cite{thoutam2015temperature,wang2015origin}.

In order to explore the origin of up-turn behavior, we replot the $\rho$$_{xx}$($\textit{T}$,$\textit{H}$) of 0~T and 7~T as well as the difference $\Delta$$\rho$$_{xx}$ = $\rho$$_{xx}$($\textit{T}$, 7~T) - $\rho$$_{xx}$($\textit{T}$, 0~T) in Fig. 2c. It clearly shows that the resistivity in a magnetic field consists of two components, $\rho_{0}(T)$ and $\Delta$$\rho$$_{xx}$, with opposite temperature dependencies. As discussed by Wang~\textit{et al.} for WTe$_{2}$ \cite{wang2015origin} , the resistivity can be written as
\begin{equation}
\rho_{xx}(T,H) = \rho_{0}(T)[1+\alpha(\frac{H}{\rho_{0}})^{m}]
\end{equation}
The second term is the magnetic-field-induced resistivity $\Delta$$\rho$$_{xx}$, which follows the Kohler's rule with two constants $\alpha$ and \textit{m}. $\Delta$$\rho$$_{xx}$ (= $\alpha$\textit{H}$^{m}$/$\rho$$_{0}$$^{m-1}$) is inversely proportional to 1/$\rho$$_{0}$ (when \textit{m} = 2) and competes with the first term upon changing temperature, possibly resulting in a minimum in $\rho(T, H)$ curves.\\

 \begin{figure}[!htbp]
   \includegraphics[width=8 cm]{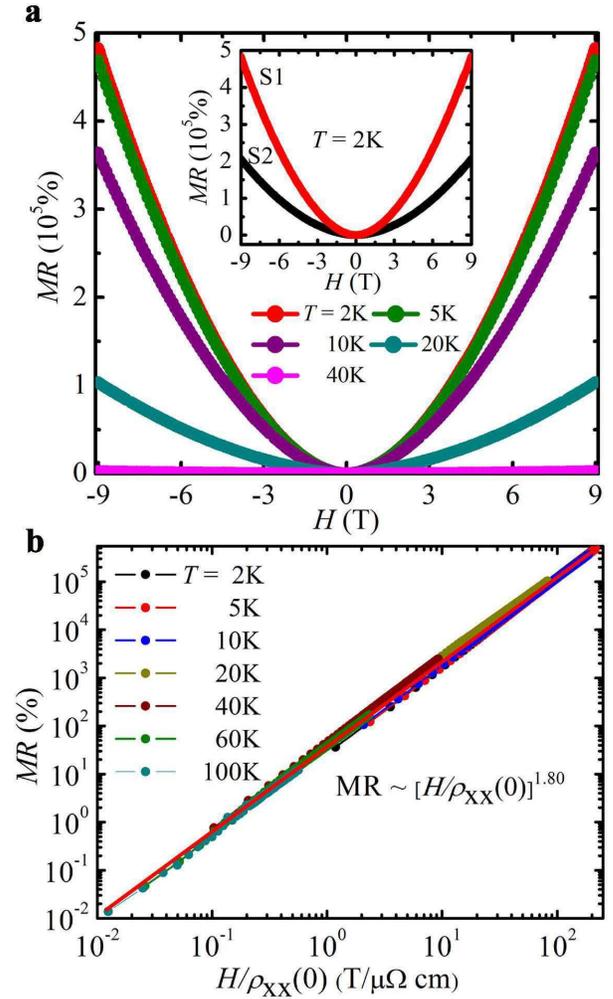}
 \caption{(Color online)  Field dependence of MR of $\alpha$-WP$_2$. (a) MR as a function of field at various temperatures. The inset compares MR as a function of magnetic field of sample 1 and sample 2.  (b) MR as a function of $\textit{H}/\rho_{xx}(0)$ plotted on log scale. The red line is the fitting using Kohler's rule scaling with \textit{m} = 1.80.}
\end{figure}
  \subsection{\textbf{\textbf{C. Longitudinal magnetoresistance}}}
 Figure 3a shows the MR as a function of field at various temperatures, with the conventional definition $\textit{MR}= \frac{\Delta\rho}{\rho(0)} = [\frac{\rho(H)-\rho(0)}{\rho(0)}]\times100\%$. The MR is extremely large at low temperatures, reaching 4.82$\times$ 10$^{5}$\% at 2 K and 9 T, and does not show any sign of saturation up to the highest field used in our measurements. The inset of Fig. 3a displays MR of sample 1 and sample 2 with different RRR values of 2491 and 1852, respectively. It is clear that the magnitudes of MR are strongly dependent on the quality of the crystals, which was also observed in Dirac semimetal PtBi$_2$ \cite{Gao2107} and $\beta$-WP$_2$ \cite{kumar2017}. As discussed above, the MR can be described by the Kohler scaling law \cite{pippard1989magnetoresistance}

\begin{equation}
\textit{MR} = \frac{\Delta\rho_{xx}(T,H)}{\rho_{0}(T)} = \alpha(\textit{H}/\rho_{0})^{m}.
\end{equation}

As shown in Fig. 3b, all MR data from T = 2~K to 100~K collapse onto a single straight line when plotted as MR $\sim \textit{H}/\rho_{0}$ curve, with  $\alpha$ = 4.5 ($\mu$$\Omega$ cm/T)$^{1.8}$ and $\textit{m}$ = 1.8 obtained by fitting. Both the same temperature dependence of MR at different fields, and the validity of Kohler scaling law at different temperatures exclude the field-induced MI transition as an origin of the up-turn behavior in $\alpha$-WP$_2$. Note that sample 2 exhibits a similar behavior, although its MR is smaller than that of sample 1 \cite{appendix} .\\

\begin{figure*}[!htbp]
\includegraphics[width= 16cm]{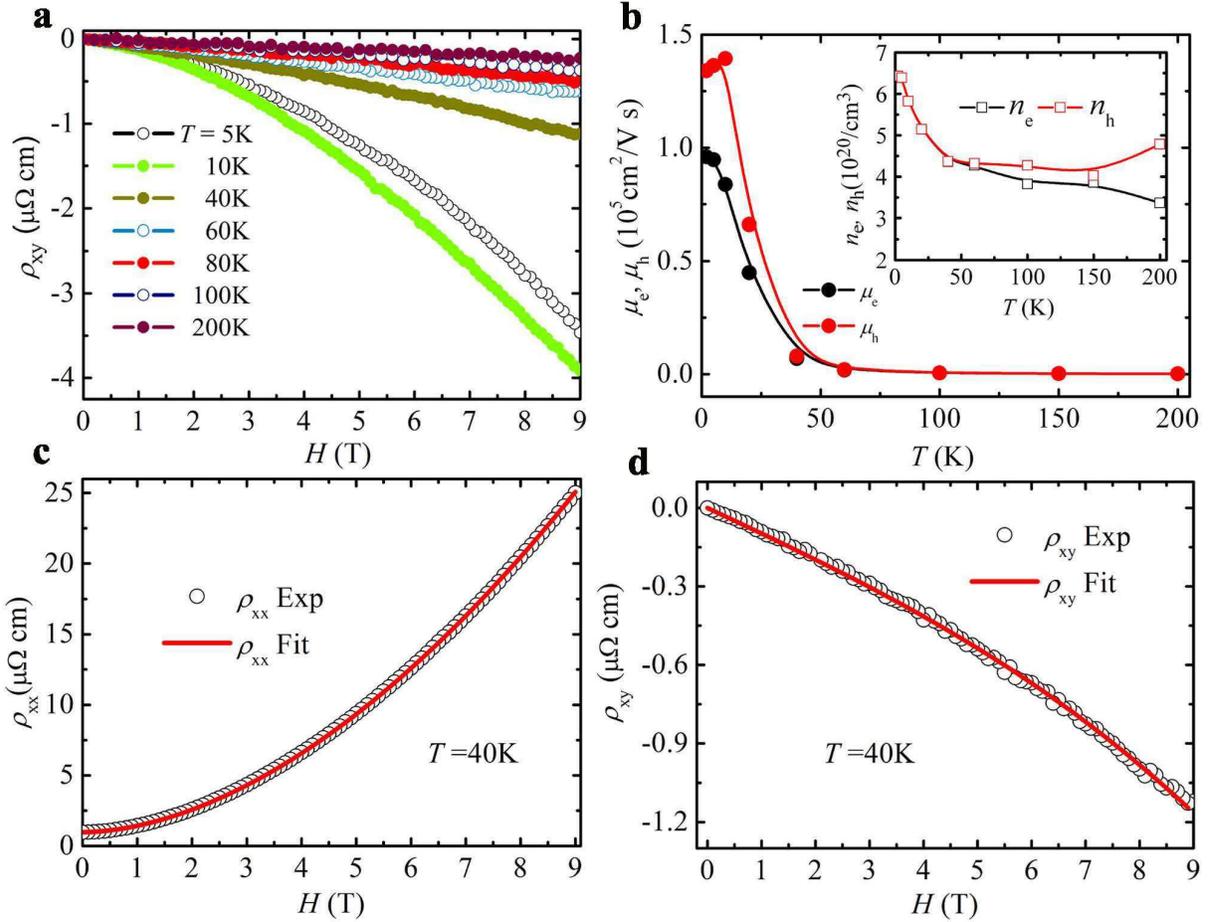}
\caption{(Color online) Charge-carriers mobility and density. (a) Field dependence of Hall resistivity $\rho$$_{xy}$ at various temperatures. (b) The mobility $\mu_{e}$ ($\mu_{h}$), the carrier density $\textit{n}_{e}$ ($\textit{n}_{h}$)(the inset) obtained by the fitting of Eq.~(3) as a function of temperature. (c) The longitudinal resistivity $\rho$$_{xx}(H)$ and (d) the Hall resistivity $\rho$$_{xy}(H)$ as a function of magnetic field, in which the red lines is the fit using the two-band model (Eq.~(3) in the text).}
\end{figure*}

\vskip 0.5cm
  \subsection{\textbf{\textbf{D. Compensation and high mobility of charge carriers}}}
According to the two-band model, the complex resistivity under an applied magnetic field \textit{H} of a semimetal is given by \cite{ali2014large}

\begin{equation}
\rho = \frac{1+\mu_{e}\mu_{h}\textit{H}^{2}+\textit{i}(\mu_{h}-\mu_{e})\textit{H}}{\textit{e}[\textit{n}_{e}\mu_{e}-\textit{n}_{h}\mu_{h}+\textit{i}(\textit{n}_{e}-\textit{n}_{h})\mu_{e}\mu_{h}\textit{H}]},
\end{equation}
where $\textit{n}_{e}$ ($\textit{n}_{h}$) is the charge density of electrons (holes), $\mu_{e}$ ($\mu_{h}$) the mobility of electrons (holes) and \textit{e} is the charge of electron. The experimentally observed longitudinal resistivity $\rho_{xx}(H)$ is given by the real part of Eq.~(3), and the Hall (transversal) resistivity $\rho_{xy}(H)$ corresponds to its imaginary part. In order to obtain the mobility and density of charge carriers, we measured the Hall resistivity $\rho_{xy}(H)$ at various temperatures, as shown in Fig. 4a. Then, using Eq.~(3) to fit directly both $\rho_{xx}(H)$ and $\rho_{xy}(H)$ data, the $\textit{n}_{e}$ ($\textit{n}_{h}$) and $\mu_{e}$ ($\mu_{h}$) values were obtained at different temperatures. Figures~4c and 4d display the $\rho_{xx}(H)$ and $\rho_{xy}(H)$ data, as well as their fits, at 40~K as a representative temperature. The data and its fits at other temperatures are shown in the Appendix \cite{appendix}.  The obtained values of $\mu_{e}$ ($\mu_{h}$) and $\textit{n}_{e}$ ($\textit{n}_{h}$) as a function of temperature are plotted in Fig. 4b and its inset, respectively. It is clear that at temperatures below 50~K $\textit{n}_{e}$ and $\textit{n}_{h}$  are practically equal, {\it e.g.} $n_e$ = $6.4275\times10^{20}$~cm$^{-3}$ and $n_h$ =$6.4285\times10^{20}$ cm$^{-3}$ at $T = 2K$, which implies the compensation of electron and hole charge carriers indeed takes place in our $\alpha$-WP$_2$ samples, similar to that discovered in WTe$_{2}$ \cite{ali2014large} and in $\beta$-WP$_2$ \cite{kumar2017}. At higher temperatures $\textit{n}_{e}$ and $\textit{n}_{h}$ start deviating, {\it e.g.}  $\textit{n}_{e}$ = 3.82 $\times 10^{20} $cm$^{-3}$ and $\textit{n}_{h}$ = 4.27 $\times 10^{20}$cm$^{-3}$ at 100~K breaking the charge-carrier compensation. Although the Kohler's rule was originally developed to account for the MR in metals, it can be derived from the two-band model (Eq.~(3)) for perfectly compensated systems as discussed by Wang \textit{et al.} for the WTe$_{2}$ compound\cite{wang2015origin} . For our $\alpha$-WP$_2$ samples, the compensation effect at low temperatures makes the Kohler's rule applicable, as discussed above. Furthermore, it was found that the charge-carrier mobilities $\mu_{e}$ ($\mu_{h}$) are enhanced at low temperatures (below 50~K), {\it e.g.}  $\mu_{e} = 9.6 \times 10^{4}$ cm$^{2}$/V~s and $\mu_{h}$ = 1.3 $\times 10^{5}$ cm$^{2}$/V~s at 2 K, which are comparable with that in WTe$_{2}$ \cite{luo2015hall} and $\beta$-WP$_2$ \cite{kumar2017}. At higher temperatures, both $\mu_{e}$ and $\mu_{h}$ exhibit an obvious decrease due to enhanced phonon thermal scattering. These results indicate that the up-turn behavior in our $\alpha$-WP$_2$ samples likely originates from the strong temperature dependence of the charge-carrier mobilities.

\begin{figure*}[!htbp]
\includegraphics[width= 16cm]{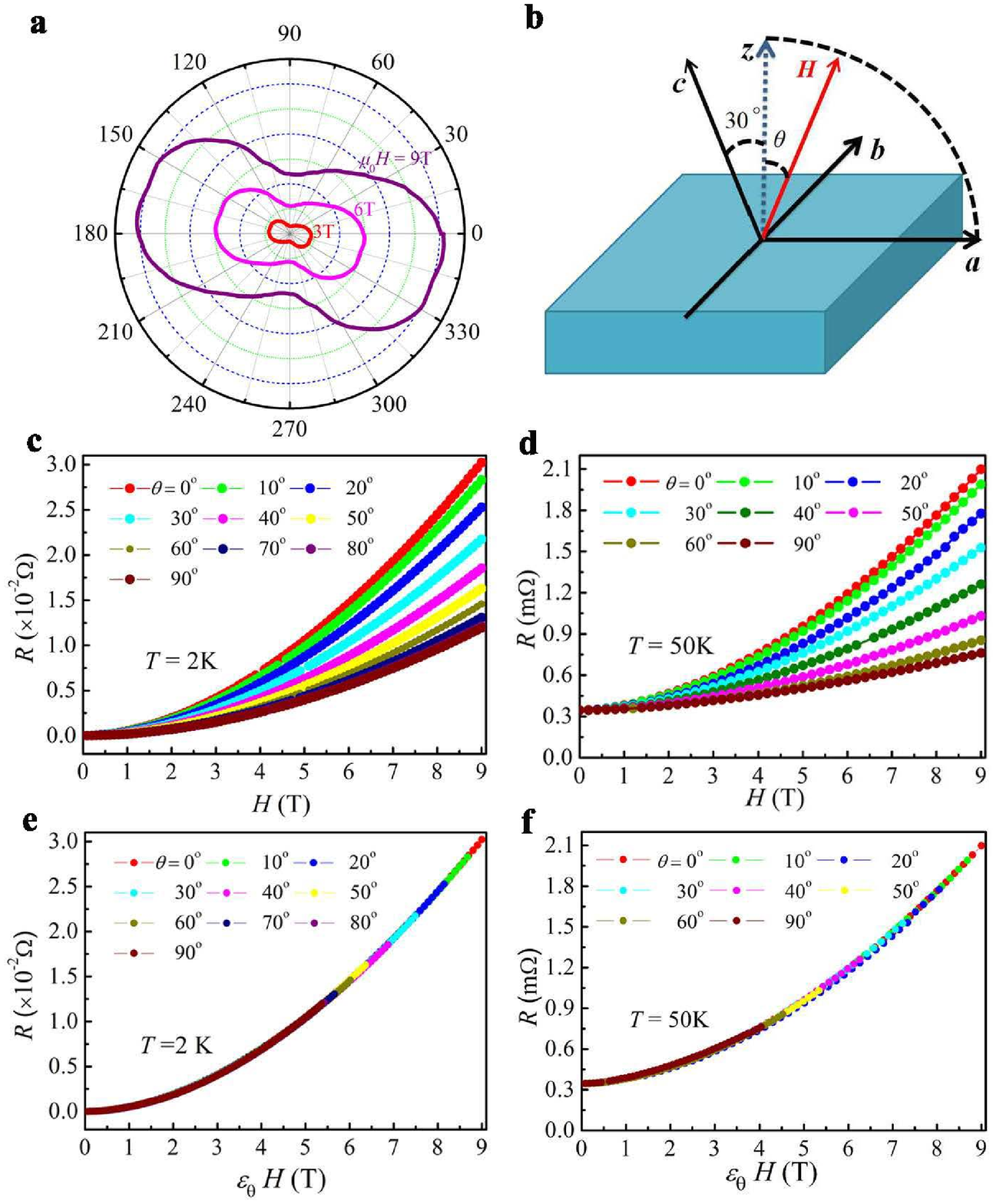}
\caption{(Color online)  Anisotropy and scaling behavior of the resistance at various magnetic field orientations. (a) Polar plot of resistivity as a function of $\theta$, magnetic field angle with respect to the \textit{z} axis. (b) Definition of the current and magnetic field directions. Current is applied along the \textit{b} axis, while the magnetic field angle $\theta$ is given with respect to the normal direction of \textit{ab} plane (c),(d) Resistance as a function of magnetic field measured at various magnetic field angles at 2 K and 50 K, respectively. (e),(f) Resistance replotted as a function of $\varepsilon_{\theta}\textit{H}$, where $\varepsilon_{\theta}$ is a scaling factor (cos$^{2}$$\theta$+$\gamma$$^{-2}$sin$^{2}$$\theta$)$^{1/2}$. }
\end{figure*}
  \subsection{\textbf{\textbf{E. The anisotropy of the resistance}}}
As discussed in Ref.~\onlinecite{thoutam2015temperature} for WTe$_{2}$, the anisotropy of the resistance reflects the Fermi surface topology. In order to address the  Fermi surface anisotropy and check whether the electronic structure changes with temperature in $\alpha$-WP$_2$ crystals, we measured the longitudinal resistance $\textit{R}_{xx}(\textit{H}, \theta)$ at a fixed temperature, where $\theta$ is the angle of applied magnetic field with respect to the \textit{z} axis which is perpendicular to \textit{ab} plane (see Fig.~5b for the definition of $\theta$). First, we measured $\textit{R}_{xx}(\theta)$ by scanning $\theta$ at 2 K under fixed magnetic fields  $\mu_{0}\textit{H}$ = 3, 6 and 9 T, respectively, as shown in Fig. 5a. The polar plot of $\textit{R}_{xx}(\theta)$ reflects the symmetry of the projected profile of the Fermi surface onto the plane perpendicular to current. The system and hence the Fermi surfaces have inversion and C$_{2x}$ symmetries. When current is applied along the \textit{x} axis, the C$_{2x}$ symmetry results in $\textit{R}_{xx}(\theta)=\textit{R}_{xx}(\pi+\theta)$ (see Fig. 5a). The data clearly reveals that the resistance is anisotropic, with largest resistance for magnetic field applied at $\theta = -15^{\circ}$ with respect to the \textit{z} axis that is 30$^{\circ}$ relative to the \textit{c} axis due to the monoclinic crystal structure of $\alpha$-WP$_2$. The minimum of resistance is close to $\theta = 90^{\circ}$, and the polar diagram has a peanut shape. The anisotropy of resistance relates to the anisotropy of Fermi surface.

Then, we measured  $\textit{R}_{xx}(\textit{H})$ at $T = 2$~K and 50~K at a fixed $\theta$, as shown in Figs. 5c and 5d, respectively. The resistance anisotropy is consistent with $\textit{R}_{xx}(\theta)$ mentioned above, with larger resistance for a fixed magnetic field applied close to \textit{z} axis ($\theta = 0^{\circ}$) for both temperatures. At the same time, we found that $\textit{R}_{xx}(\textit{H})$ curves obtained at a fixed temperature but at various angles $\theta$ can be collapsed onto a single curve with a field scaling factor $\varepsilon$$_\theta$ = (cos$^{2}$$\theta$+$\gamma$$^{-2}$sin$^{2}$$\theta$)$^{1/2}$, as shown in Figs.~5e and 5f, where $\gamma$ is a constant at a given temperature. That is, the resistance of $\alpha$-WP$_2$ has a scaling behavior $\textit{R}_{xx}(\textit{H}, \theta)$ = $\textit{R}_{xx}(\varepsilon_{\theta}\textit{H})$, where $\varepsilon_{\theta}\textit{H}$ is the reduced magnetic field and $\varepsilon$$_\theta$ = (cos$^{2}$$\theta$+$\gamma$$^{-2}$sin$^{2}$$\theta$)$^{1/2}$ reflects the mass anisotropy for an elliptical Fermi surface, with $\gamma$ being the ratio of the effective masses of electrons moving in directions given by $\theta = 0^{\circ}$ and  90$^{\circ}$. This anisotropic scaling rule has also used to account for the angular dependence of MR in graphite \cite{soule1958magnetic,noto1975simple}, WTe$_{2}$ \cite{thoutam2015temperature}, as well as the anisotropic properties of high temperature superconductors \cite{ishida1996anisotropy,blatter1992isotropic}. For our $\alpha$-WP$_2$ sample, we obtained $\gamma$ = 1.68, 1.72, 2.57, 2.23 and 1.98 from $\textit{R}_{xx}(\textit{H})$ data at \textit{T} = 2.0, 10, 30, 50 and 100 K, respectively. The data and fits at other temperatures are shown in the Appendix \cite{appendix}. We also measured $R_{xx}(H,\phi)$, where $\phi$ is the angle of $H$ with respect to the $z$ axis, but within the $bz$ plane, current was also applied along the $b$ axis \cite{appendix}. No negative magnetoresistance was observed for $H$ parallel to the current direction.
  \subsection{\textbf{\textbf{F. Discussion}}}
The investigated $\alpha$-WP$_2$ with monoclinic crystal structure (C2/m) is a topologically trivial semimetal as predicted by our band calculations, which contrasts this material to its $\beta$-phase polymorph that has noncentrosymmetric orthorhombic crystal structure ($Cmc2_{1}$) and was predicted to be a robust type-II Weyl semimetal \cite{autes2016robust}. The comparison of MR behaviors and electronic structure of $\alpha$-WP$_2$ and $\beta$-WP$_2$ may provide some hints on the XMR mechanism as both materials have the same composition and very similar Fermi surfaces \cite{appendix}. From the above experimental results, $\alpha$-WP$_2$ exhibits all typical characteristics of XMR materials, such as the nearly quadratic field dependence of MR and the field-induced up-turn in resistivity following by a plateau at low temperatures. Phenomenologically, these properties can be understood from the point of view of compensation of electron and hole charge carriers with  high mobilities at low temperatures, which was confirmed by our Hall measurements and band structure calculations. Another similarity with $\beta$-WP$_2$ is the strong dependence of the magnitudes of MR on the RRR value, {\it i.e.} the sample quality. The robustness of Weyl semimetal phase in $\beta$-WP$_2$ due to the same chirality of the neighboring Weyl nodes is believed to suppress the backscattering \cite{kumar2017}, resulting in small $\rho_0$ and large MR. However, there are two differences in MR behaviors between $\alpha$-WP$_2$ and $\beta$-WP$_2$. The first difference is related to the anisotropy of MR. In $\beta$-WP$_2$, MR reaches its maximum and minimum when $H$ is parallel to the $b$- and $c$-axis, respectively, while current is applied along the $a$-axis (easy growth axis), as shown in Fig.~3e of Ref.~\onlinecite{kumar2017}. $\beta$-WP$_2$ exhibits a strong MR anisotropy, much more pronounced compared to WTe$_2$. Kumar {\it et al.} suggested that such strong anisotropy in MR is due to the shape of spaghetti-type open FSs \cite{kumar2017}, {\it i.e.} when the field is parallel to the $c$-axis, the perpendicular cross-section area of FS becomes open which would result in a dramatic drop of MR. The lack of inversion symmetry in $\beta$-WP$_2$ leads to spin-splitting of bands. The hole FS pockets are open extending along the $b$-axis, while electrons form a pair of bowtie-like closed pockets \cite{appendix}. In contrast,  for $\alpha$-WP$_2$ MR reaches its maximum when $H$ is oriented along the direction at 15$^\circ$ with respect to the $c$-axis, and has a minimum at $H$ parallel to the $a$-axis, with current is applied along the $b$-axis (easy growth axis). In this case, the polar diagram has a peanut-like shape as shown in Fig.~5a. MR in $\alpha$-WP$_2$ exhibits a weaker anisotropy compared to that of $\beta$-WP$_2$ and the maximum of MR does not occur when $H$ is applied along $c$-axis, but rather at an angle to it. We believe that the MR anisotropy in $\alpha$-WP$_2$ is also related to the topology of FSs, as shown in Fig. 1c, opening electron pockets without band spin-splitting due to the presence inversion symmetry. The second difference concerns the validity of Kohler's scaling law. In $\alpha$-WP$_2$, the MR data in a wide temperature range can be described well by this law, while the MR of $\beta$-WP$_2$ above 10~K deviates from Kohler's rule considerably \cite{wang2017large}. This indicates that the temperature-induced Lifshitz transition as a possible origin of XMR mechanism cannot be ruled out in $\beta$-WP$_2$, as also suggested for WTe$_2$ \cite{wu2015temperature}.

It is also interesting to conduct a comparison with the MR behavior of WTe$_2$. A remarkable difference between the investigated $\alpha$-WP$_2$ and WTe$_{2}$ is that the change of $\gamma$ value ($\sim$2.0) with temperature is not obvious in our $\alpha$-WP$_2$ samples, while in WTe$_{2}$ at low temperatures $\gamma$ is almost 2.5 times higher than that at high temperatures \cite{thoutam2015temperature}. In fact, Wu \textit{et al.} confirmed the existence of a temperature-induced Lifshitz transition ({\it i.e.} the complete disappearance of hole pockets at high temperatures)  in WTe$_{2}$ by means of ARPES and thermoelectric power measurements \cite{wu2015temperature}, which is believed to be the origin of up-turn behavior in this material. However, the absence of obvious temperature dependence of $\gamma$ in $\alpha$-WP$_2$ indicates that the temperature-induced Fermi surface transition should not be the origin of up-turn behavior in this case.
 \section{\textbf{IV. CONCLUDING REMARKS}}
In summary, we successfully synthesized $\alpha$-WP$_2$ crystals and performed their magneto-transport measurements and electronic structure investigation. It was found that $\alpha$-WP$_2$ exhibits practically all typical characteristics of XMR materials. Our Hall resistivity measurements and band structure calculations reveal that the compensation effect and high mobility of carriers take place in $\alpha$-WP$_2$. The facts that the normalized MR under different magnetic fields has the same temperature dependence in $\alpha$-WP$_2$, the Kohler scaling law can describe the MR data in a wide temperature range, and the independence of anisotropic parameter $\gamma$ on temperature rule out both field-induced gap and the temperature-induced Lifshitz transition as the origins of up-turn behaviors in $\alpha$-WP$_2$ semimetal. We also found that the resistance polar diagrams has a peanut shape when magnetic field is rotated in the $\textit{ac}$ plane, which can be understood by the open-orbit electrons pockets in FS. However, the mechanism underlying the sharp enhancement of $\mu_{e}$  and $\mu_{h}$ at low temperatures remains to be addressed. Our findings highlight $\alpha$-WP$_2$ as a new material for exploring XMR phenomena.
 \section{\textbf{ACKNOWLEDGMENTS}}
We are grateful for support from the National Basic Research program of China under Grant No. 2016YFA0300402, 2015CB921004 and the National Natural Science Foundation of China (No. 11374261), the Zhejiang Provincial Natural Science Foundation (No. LY16A040012) and NCCR Marvel. Q.W. and O.V.Y. acknowledge support by the NCCR Marvel. First-principles calculations were performed at the Swiss National Supercomputing Centre (CSCS) under project s675 and the facilities of Scientific IT and Application Support Center of EPFL.


 \section{\textbf{APPENDIX}}
  \subsection{Comparison of Fermi surfaces of $\alpha$-WP$_2$ and $\beta$-WP$_2$}
\begin{figure*}[!htbp]
\includegraphics[width= 13cm]{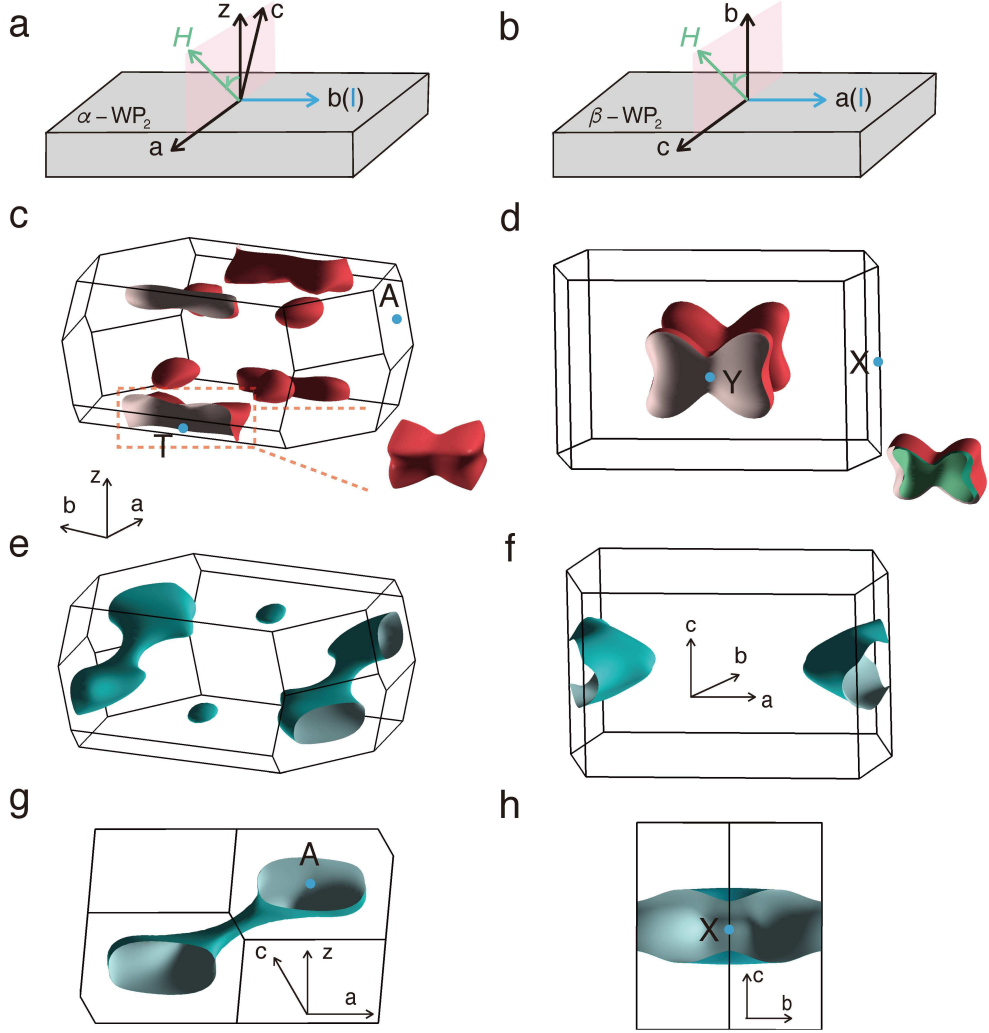}
\caption{(Color online) (a),(b) Definitions of the crystal axes, magnetic field and current directions used for $\alpha$-WP$_2$ and $\beta$-WP$_2$, respectively. (c),(d) Electron Fermi surfaces for $\alpha$-WP$_2$ and $\beta$-WP$_2$, respectively. The entire deformed bowtie-shaped pocket in $\alpha$-WP$_2$ and the pair of spin-split nested Fermi surfaces in $\beta$-WP$_2$ are shown separately (bottom right of panels (c) and (d), respectively). (e),(f) Hole Fermi surfaces of $\alpha$-WP$_2$ and $\beta$-WP$_2$, respectively. (g),(h) Open hole Fermi surfaces of $\alpha$-WP$_2$ and $\beta$-WP$_2$, respectively, shown along the direction they extend. }
\label{fig6}
\end{figure*}

The as-grown crystal of the two phases of WP$_2$ both have one distinct long dimension. The investigated $\alpha$-WP$_2$ crystal is of size 0.7$\times$0.2$\times$0.1 mm$^3$ with its longest dimension (0.7~mm) aligned along axis $b$. Likewise, single crystals of  $\beta$-WP$_2$ reported in Ref. \onlinecite{kumar2017} are needle-shaped with longer dimension aligned along the $a$ axis. This characteristic shape makes it difficult to apply current along other shorter axes. Further, considering the similarity of the Fermi surfaces of these two phase, it is more convenient to make the following correspondance between the axes of their crystal structure, $a$ axis of $\alpha$-WP$_2$ corresponds to $b$ axis of $\beta$-WP$_2$,  $b$ axis of $\alpha$-WP$_2$ to $a$ axis of $\beta$-WP$_2$, and $c$ axis of $\alpha$-WP$_2$ to $c$ axis of $\beta$-WP$_2$. Furthermore, we note that $\alpha$-WP$_2$ has monoclinic crystal structure, i.e. $c$ axis is not orthogonal to the $ab$ plane, while $\beta$-WP$_2$ has orthorhombic crystal structure, as shown in Figs.~\ref{fig6}a,b of this Appendix.

Figures~\ref{fig6}c-f compare the Fermi surfaces of $\alpha$-WP$_2$ and $\beta$-WP$_2$. For convenience, the electron and hole pockets are plotted individually in each case. Both materials have closed electron Fermi surfaces of similar bowtie shape, although in the case of $\alpha$-WP$_2$ they appear to be more deformed, as well as open tube-shaped hole Fermi surfaces. In the case of $\alpha$-WP$_2$, the deformed bowtie closed electron Fermi surface is located at the T point of the Brillouin zone (Fig.~\ref{fig6}c), while the warped tube-shaped open hole Fermi surface encloses the A point (Fig.~\ref{fig6}e) and extends along the direction perpendicular to the $c$ axis in $ac$ plane (Fig.~\ref{fig6}g). In the case of $\beta$-WP$_2$, the bowtie-like closed electron Fermi surface is located at the Y point of the Brillouin zone (Fig~\ref{fig6}d), and the tube-shaped open hole Fermi surface encloses the X point (Fig.~\ref{fig6}f) extending along the $a$ axis (Fig.~\ref{fig6}h). Hence, in both materials the open orbits extend along the direction normal to the $c$ axis in $ac$ and $bc$ planes, respectively. However, there is a difference between the two Fermi surfaces--small pockets in both electron and hole Fermi surface of $\alpha$-WP$_2$ can be seen (Fig.~\ref{fig6}c,e), in contrast to $\beta$-WP$_2$.

Furthermore, the crystal structure of $\alpha$-WP$_2$ belongs to centrosymmetric space group C2/m (No.~12), while that of $\beta$-WP$_2$ belongs to space group $Cmc2_1$ (No.~36) that lacks inversion symmetry. As a consequence, the Fermi surface of $\alpha$-WP$_2$ has twofold spin degeneracy, while the Fermi surface of $\beta$-WP$_2$ is composed of a pair of surfaces nested inside each other when SOC is taken into account (see the example for the electron Fermi surface in Fig.~\ref{fig6}d).

  \subsection{Topological properties of $\alpha$-WP$_2$}
\begin{figure*}[!htbp]
\includegraphics[width= 13cm]{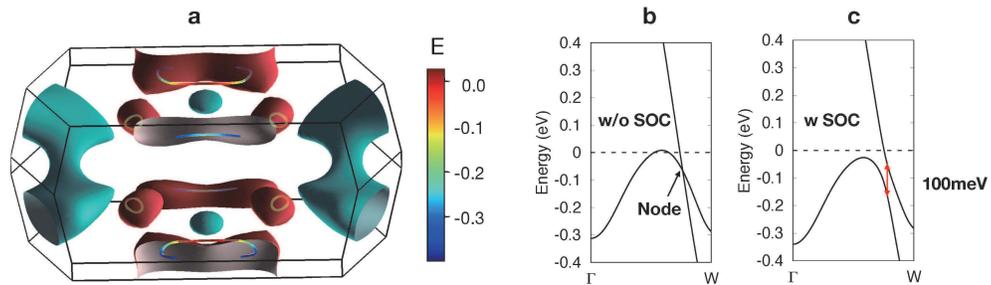}
\caption{(Color online) (a) Nodal-line degeneracy and the Fermi surface of $\alpha$-WP$_2$. (b) Band structure calculated without considering SOC along a $k$-point path that crosses the nodal point. (c) A gap of about 100~meV at the nodal point is present when SOC is taken into account.}
\label{fig-nodalline}
\end{figure*}

Since $\alpha$-WP$_2$ has PT symmetry, that is the combination of inversion and time-reversal symmetry, a nodal-line degeneracy can be present in such system \cite{PhysRevB.92.045108}. From the band structure shown in Figs.~1f,g of the main text, the band inversion character along the $\Gamma-$T direction is evident. By constructing the symmetrized Wannier tight-binding model without taking into account SOC \cite{MOSTOFI20142309}, a nodal line between bands $N$ and $N+1$ was identified with the help of WannierTools software \cite{wu2017wanniertools}, as shown in Fig.~\ref{fig-nodalline}a. The band structure plotted along a $k$-point path across the nodal point is presented in Fig.~\ref{fig-nodalline}b, showing type-II behavior character of the degeneracy. Upon taking SOC into consideration, the degeneracy is lifted by upto {\it ca.}~100~meV (Fig.~\ref{fig-nodalline}c) \cite{PhysRevB.92.045108}.

\begin{figure*}[!htbp]
\includegraphics[width= 13cm, angle=270]{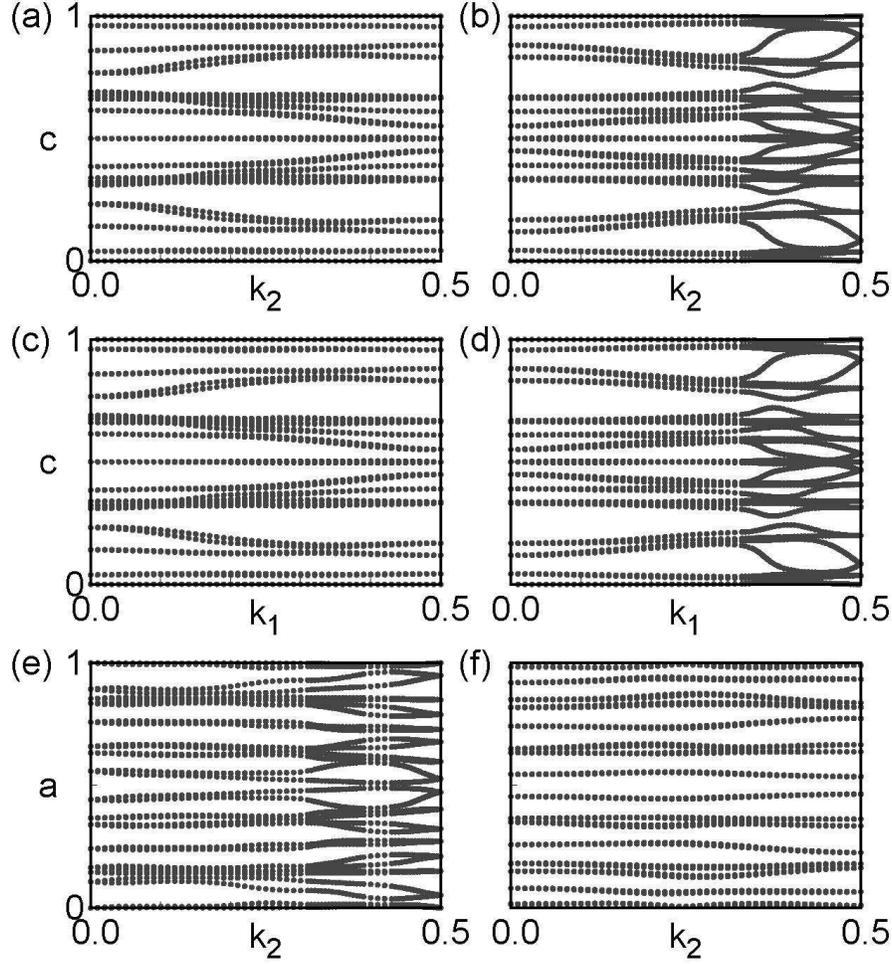}
\caption{(Color online) Wilson loops of six time-reversal invariant planes (a) $k_1$=0.0; (b) $k_1$=0.5; (c) $k_2$=0.0; (d) $k_2$=0.5; (e) $k_3$=0.0; (f) $k_3$=0.5, where $k_1$, $k_2$, $k_3$ are in units of the reciprocal lattice vectors. }
\label{wilsonloop}
\end{figure*}

In order to determine the $\mathcal{Z}_2$ indices, the Wilson loops were calculated on six time-reversal invariant planes using WannierTools \cite{wu2017wanniertools}. The results are shown in Fig.~\ref{wilsonloop}. According to the Wilson loop definition \cite{PhysRevB.84.075119, PhysRevB.83.035108}, the topological indices are (0;000) \cite{LiangFuPRL} which identifies $\alpha$-WP$_2$ as a topologically trivial semimetal when SOC is taken into account.

\begin{figure*}[!htbp]
\includegraphics[width= 16cm]{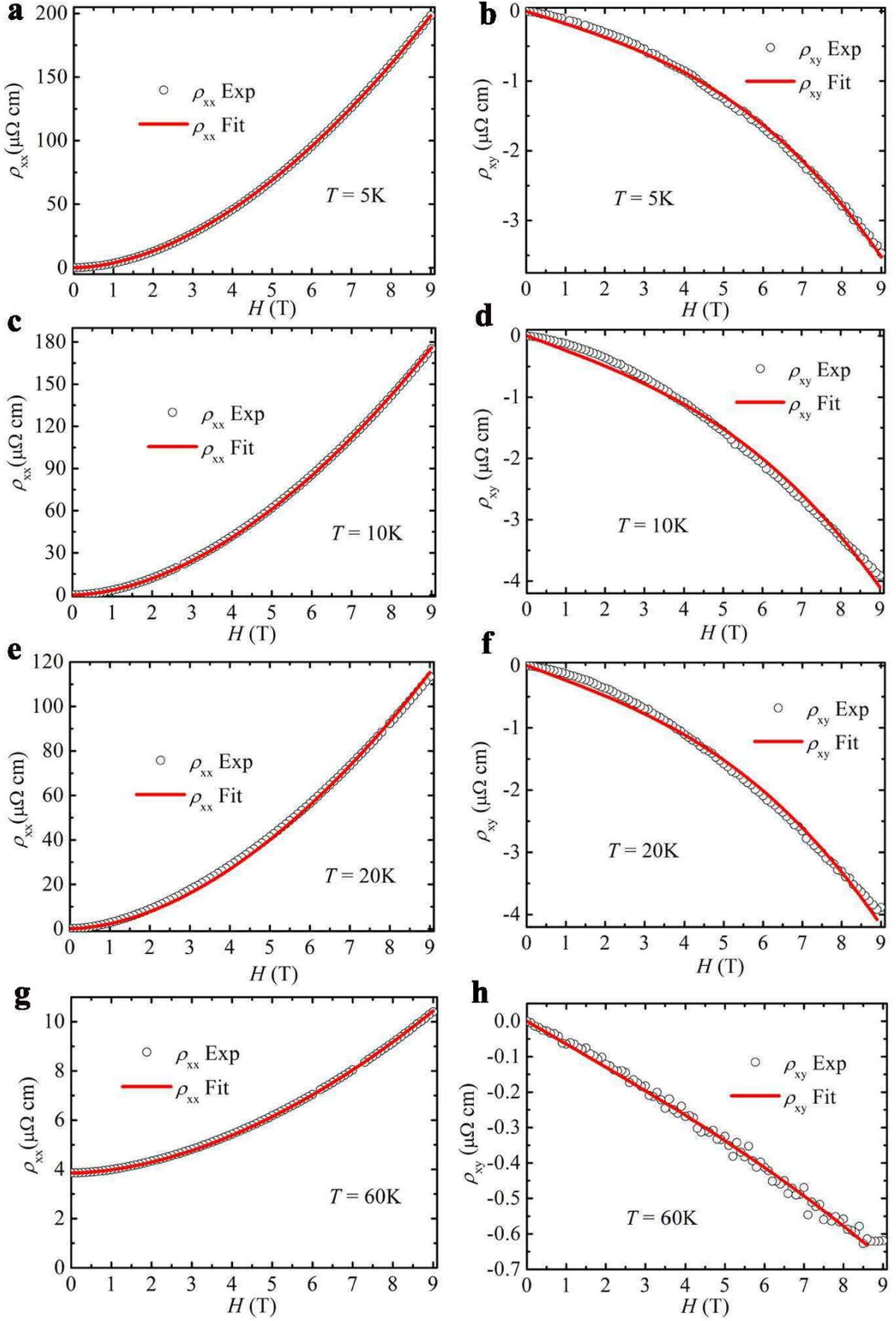}
\caption{(Color online) Longitudinal resistivity $\rho$$_{xx}(H)$ and Hall resistivity $\rho$$_{xy}(H)$ at different temperatures. (a),(c),(e),(g) Longitudinal resistivity $\rho$$_{xx}(H)$ as a function of field at 5, 10, 20 and 60K, respectively. (b),(d),(f),(g) Hall resistivity $\rho$$_{xy}(H)$ as a function of field at 5, 10, 20 and 60K, respectively.}\label{figS2}
\end{figure*}

\begin{figure*}[!htbp]
\includegraphics[width= 16cm]{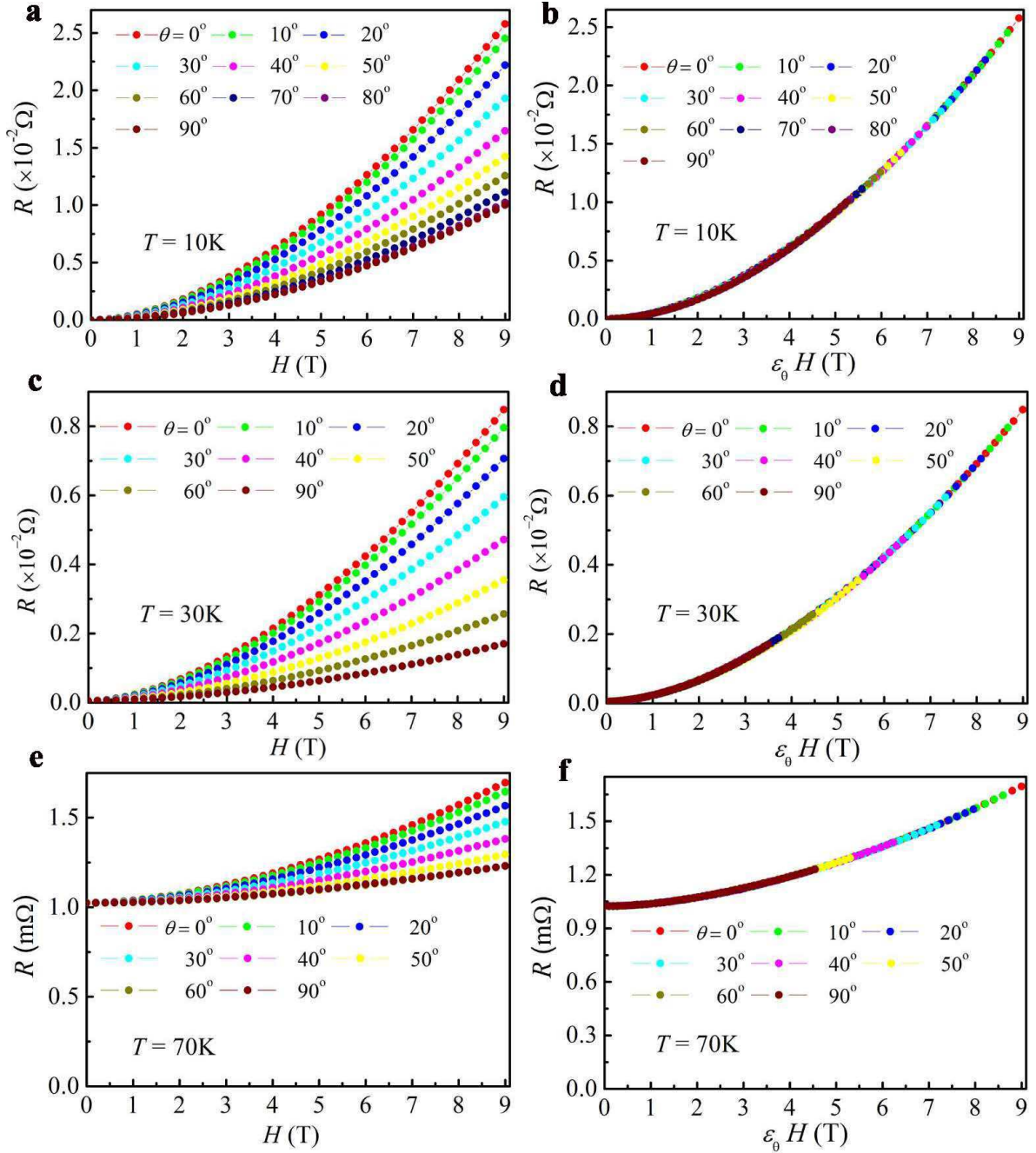}
\caption{(Color online) Resistances as a function of field for various magnetic field orientations of $\alpha$-WP$_2$ and different temperatures. (a),(c),(e) Resistance as a function of magnetic field measured at various magnetic field angles at 10 K, 30 K and 70 K, respectively. (b), (d), (f) Resistance replotted as a function of $\varepsilon_{\theta}\textit{H}$, where $\varepsilon_{\theta}$ is a scaling factor (cos$^{2}$$\theta$+$\gamma$$^{-2}$sin$^{2}$$\theta$)$^{1/2}$.}\label{figS3}
 \end{figure*}

\begin{figure*}[!htbp]
\includegraphics[width= 8cm]{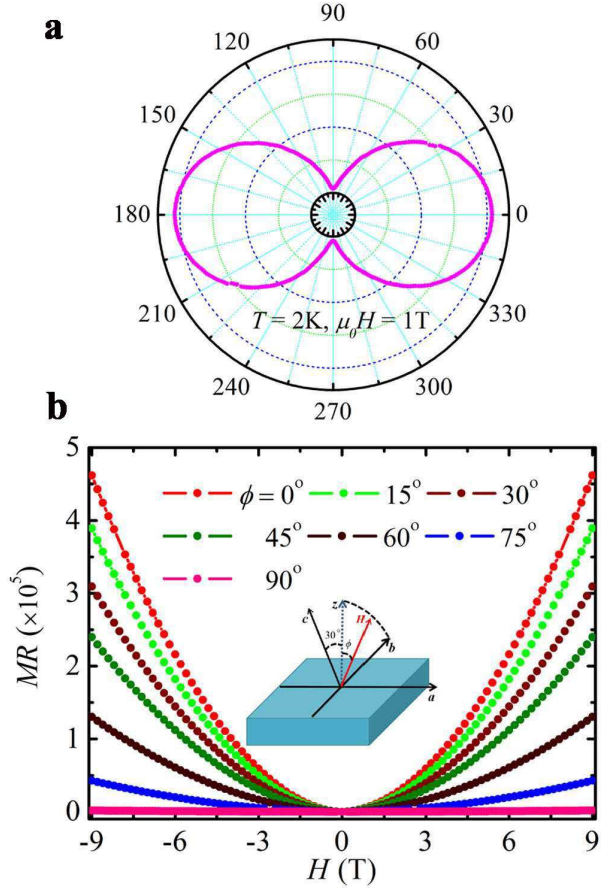}
\caption{(Color online) Field dependence of MR at various magnetic field orientations. (a) Polar plot of resistivity as a function of $\phi$, magnetic field angle with respect to the \textit{z} axis. (The inset shows the definition of the current and magnetic field directions. (b) MR of $\alpha$-WP$_2$ as a function of field at various of angles and \textit{T} = 2K.}
 \end{figure*}

\begin{figure*}[!htbp]
\includegraphics[width= 16cm]{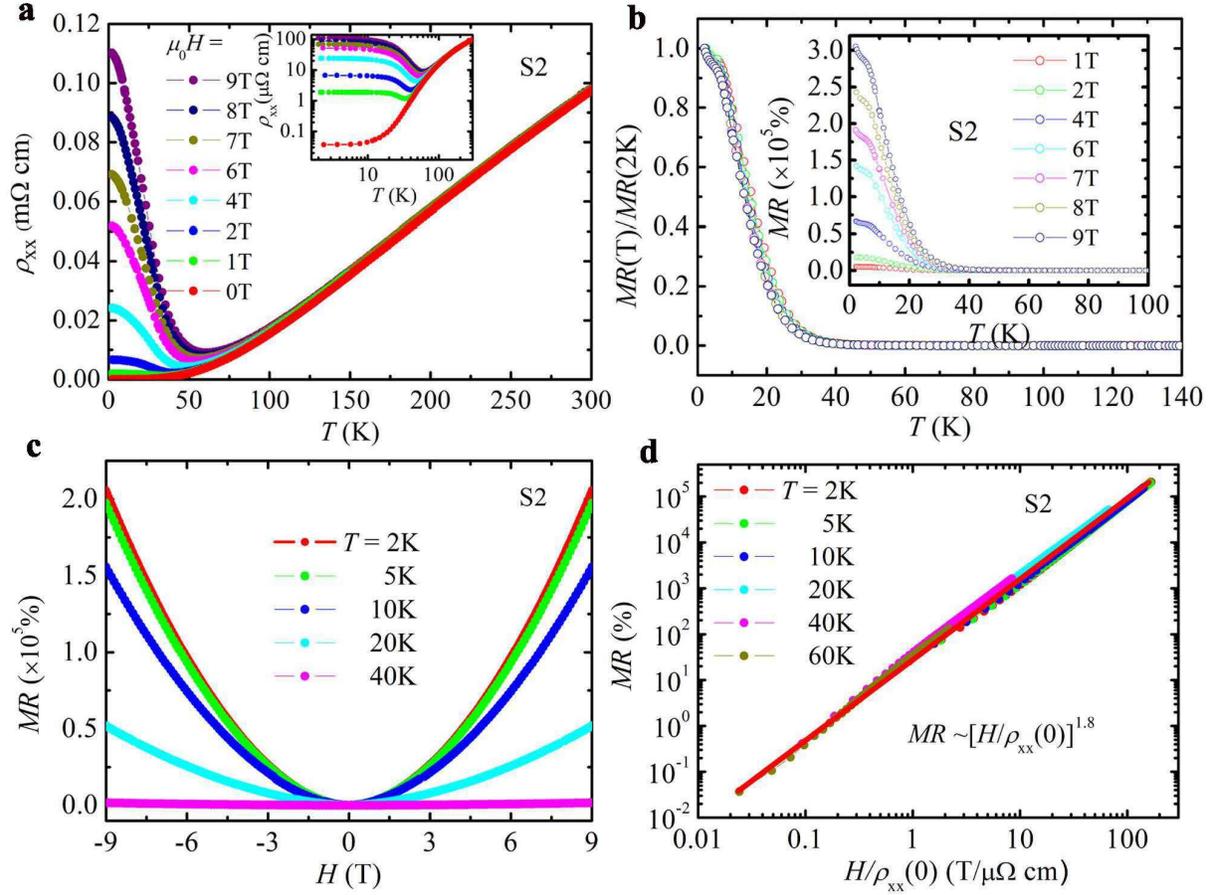}
\caption{(Color online) Resistivity and MR of sample 2. (a) Resistivity for sample 2 as a function of temperature at various magnetic fields. The inset plotted the same data using a \textit{log} scale, for showing the plateau in lower temperature. (b) Temperature dependence of the normalized MR by the MR value at 2 K at various magnetic fields of sample 2. The inset is the original data of MR as a function of temperature. (c) MR as a function of field at various temperatures. (d) MR as a function of $\textit{H}/\rho_{xx}(0)$ plotted by a log scale. The red line is the fitting using Kohler's rule scaling with \textit{m} = 1.80.}
 \end{figure*}

\bibliography{Reference}

\begin{thebibliography}{91}
\bibitem [{app()}]{appendix}%
  \BibitemOpen
  \href@noop {} {}\bibinfo {note} {See Appendix at ... for the details of
  comparison of Fermi surfaces for $\alpha$-WP$_2$ and $\beta$-WP$_2$,
  discusion for nodal lines and measurement of $R_{xx}(H,\beta)$, when $H$ is
  rotated in the $bz$ plane, current applied along $b$ axis.}\BibitemShut
  {Stop}%
\end{thebibliography}
\end{document}